\documentclass[a4paper,11pt]{article}
\usepackage{pos}
\usepackage{xspace}
\usepackage{orcidlink}
\usepackage{overpic}
\usepackage{rviewport}
\usepackage{multicol}
\usepackage{subfigure}
\usepackage{mathtools}
\usepackage{bm}
\usepackage{tikz}
\usetikzlibrary{shapes}
\usepackage{setspace}
\usepackage{upgreek}
\usepackage{wrapfig}

\newcommand{\modelNe}{\textnormal{\texttt{neCL}}\xspace}
\newcommand{\modelXr}{\textnormal{\texttt{expX}}\xspace}
\newcommand{\modelBase}{\textnormal{\texttt{base}}\xspace}
\newcommand{\modelSpur}{\textnormal{\texttt{spur}}\xspace}
\newcommand{\modelCre}{\textnormal{\texttt{cre10}}\xspace}
\newcommand{\modelSyn}{\textnormal{\texttt{synCG}}\xspace}
\newcommand{\modelTwist}{\textnormal{\texttt{twistX}}\xspace}
\newcommand{\modelKappa}{\textnormal{\texttt{nebCor}}\xspace}
\newcommand{\modelBub}{\textnormal{\texttt{locBub}}\xspace}
\newcommand{\modelAsymT}{\textnormal{\texttt{asymT}}\xspace}
\newcommand{\JF}{\textnormal{\texttt{JF12}}\xspace}
\newcommand{\UF}{\textnormal{\texttt{UF23}}\xspace}
\newcommand{\KST}{\textnormal{\texttt{KST24}}\xspace}

\newcommand{\fsz}{\ensuremath{z_\text{d}}\xspace}
\newcommand{\fswz}{\ensuremath{w_\text{d}}\xspace}

\newcommand{\pitch}{\ensuremath{\alpha}\xspace}

\newcommand{\BNT}{\ensuremath{B_\text{N}}\xspace}
\newcommand{\BST}{\ensuremath{B_\text{S}}\xspace}
\newcommand{\zT}{\ensuremath{z_\text{t}}\xspace}
\newcommand{\rT}{\ensuremath{r_\text{t}}\xspace}
\newcommand{\rNT}{\ensuremath{r_\text{tN}}\xspace}
\newcommand{\rST}{\ensuremath{r_\text{tS}}\xspace}
\newcommand{\wT}{\ensuremath{w_\text{t}}\xspace}
\newcommand{\BP}{\ensuremath{B_\text{p}}\xspace}
\newcommand{\zP}{\ensuremath{z_\text{p}}\xspace}
\newcommand{\rP}{\ensuremath{r_\text{p}}\xspace}
\newcommand{\wP}{\ensuremath{w_\text{p}}\xspace}

\newcommand{\PI}{\ensuremath{\text{PI}}\xspace}
\newcommand{\RM}{\ensuremath{\text{RM}}\xspace}
\newcommand{\RMs}{{\RM}s\xspace}
\DeclareMathAlphabet{\mathcal}{OMS}{cmsy}{m}{n}
\newcommand{\R}{\ensuremath{{\mathcal R}\xspace}}
\newcommand{\HEALPix}{{\scshape HEALPix}\xspace}

\makeatletter
\newcommand\mathcircled[1]{%
  \mathpalette\@mathcircled{#1}%
}
\newcommand\@mathcircled[2]{%
  \tikz[baseline=(math.base)] \node[draw,ellipse,inner sep=1pt] (math) {$\m@th#1#2$};%
}
\makeatother

\title{The Galactic Magnetic Field and UHECR Deflections}

\author*[a]{Michael Unger \orcidlink{0000-0002-7651-0272}\,}
\author[b]{Glennys R.\ Farrar \orcidlink{0000-0003-2417-5975}\,}

\affiliation[a]{Institute for Astroparticle Physics, Karlsruher Institute of Technology, Karlsruhe 76344, Germany}

\affiliation[b]{Center for Cosmology and Particle Physics, Department of Physics, New York University, NY 10003, USA}

\emailAdd{michael.unger@kit.edu}
\emailAdd{gf25@nyu.edu}

\abstract{ Ultrahigh-energy cosmic rays (UHECRs) experience
  deflections as they traverse the Galactic magnetic field (GMF),
  which must be accounted for when tracing them back to their sources.
  After briefly summarizing our results on uncertainties in cosmic-ray
  deflections from the UF23 ensemble of GMF models (Unger \& Farrar,
  2024), we report a new preliminary fit of the GMF including
  foreground emission from the Local Bubble. This fit uses the analytic model
  of Pelgrims et al.\ (2024) for the magnetic field in the thick shell of
  Galactic bubbles. We also discuss how variations in toroidal halo field
  modeling account for the key differences between the Jansson
  \& Farrar (2012) GMF model and the UF23 ensemble.

Furthermore, we extend our previous analysis of the origin of the
highest-energy “Amaterasu” event observed by the Telescope Array to
include the four highest-energy events detected by the Pierre Auger
Observatory. Amaterasu and PAO070114 are the UHECR events with the
smallest localization uncertainties of 4.7\% and 2.4\%,
respectively. Neither of their back-tracked directions aligns with
any compelling candidate for a continuous UHECR accelerator. This
strengthens the evidence that at least a fraction of the highest
energy events originate from transient sources.  }

\FullConference{7th International Symposium on Ultra High Energy Cosmic Rays (UHECR2024)\\
 17-21 November 2024\\
Malargüe, Mendoza, Argentina\\}

\begin{document}
\maketitle

\section{The Coherent Magnetic Field of the Milky Way}

A good understanding of the the coherent magnetic field of the Galaxy
(GMF) is important to study the impact of magnetic deflections on
the arrival directions of extragalactic ultrahigh-energy cosmic rays.
Determining the large-scale structure of the magnetic field
of our Galaxy is particularly challenging since one must infer it from
from the vantage point of Earth located inside the
field.
As demonstrated by the 2012 Jansson Farrar (\JF)
model~\cite{Jansson:2012pc}, the global structure of the GMF can be
derived by fitting suitable parametric descriptions of the structure
of the disk, toroidal and poloidal field, to the two astrophysical data
sets that provide the strongest constraints on the coherent magnetic fields:
the {\itshape rotation measures} ({\RM}s) of extragalactic polarized
radio sources and the {\itshape polarized intensity} (\PI) of the
synchrotron emission of cosmic-ray electrons in the Galaxy. The
interpretation of this data relies on {\itshape auxiliary models} of
the three-dimensional density of thermal electrons $n_e$ and cosmic-ray
electrons $n_\text{cre}$ in the Galaxy. In Ref.~\cite{Unger:2023lob} (\UF in the
following) we studied the combination of different data sets,
auxiliary models and parametric functions and obtained an ensemble of
GMF models that reflect the uncertainties and degeneracies inherent in
the inference of the global field structure from the limited
information provided by \RM and \PI data. We have narrowed down
these model variants to a few benchmark models that encompass the
largest differences within the ensemble as summarized in
Table~\ref{tab:models}.

In addition to the ensemble of eight models in \UF we introduce here a
ninth model called \modelBub, in which we include the foreground synchrotron
emission from the Local Bubble into the GMF fit, and a model called \modelAsymT discussed below. The local bubble is a
cavity with a low density of gas within which the Sun is situated and
which is thought to originate from multiple supernova explosions that
occurred in the past 10-15~Myr. A first fit of the GMF including the
Local Bubble
\begin{wraptable}{r}{0.58\textwidth}
  \vspace*{-0.2cm}
  \caption{Overview of model variations discussed in this paper. The first eight rows are the \UF model variations, and the last two rows refer to models discussed in these proceedings. \label{tab:models} \vspace*{-0.2cm}}
 \scriptsize
\vspace*{-0.cm}  \begin{tabular}{llc}
name   &  variation & $\chi^2/{\rm ndf}$ \\ \hline
\modelBase  & fiducial model &  $\text{1.22}$  \\
\modelXr   & radial dependence of X-field & $\text{1.30}$   \\
\modelSpur & replace grand spiral by local spur (Orion arm)& $\text{1.23}$   \\
\modelNe   & change $n_e$ from YMW16 t o NE2001 &   $\text{1.19}$   \\
\modelTwist & unified halo model via twisted X-field  &   $\text{1.26}$      \\
\modelKappa & $n_e$-$B$ correlation &   $\text{1.22}$   \\
\modelCre & $n_\text{cre}$ vertical scale height & $\text{1.22}$   \\
\modelSyn  & use {\scshape CosmoGlobe} synchrotron maps    & $\text{1.50}$ \\ \hline
\modelAsymT & as \modelNe, but forced N-S-asymmetric toroidal halo& $\text{1.22}$\\
\modelBub & local bubble (preliminary, spherical approximation)& $\text{1.17}$\\
\hline \vspace*{-0.5cm}
  \end{tabular}
\end{wraptable}
was recently reported in Ref.~\cite{Korochkin:2024yit} using an approximate functional form
for a magnetic field compressed into a spherical shell. Here, we present a preliminary fit of
the Local Bubble with the divergence-free analytic ``SCO'' model of
Ref.~\cite{Pelgrims:2024jko}.

As shown in Table~\ref{tab:models}, this new preliminary \modelBub fit
results in a slightly better fit quality than any of the other model
variants. The latitude of the direction of the fitted pre-explosion
field ($B_0 = 1.7~\upmu$G) agrees well with the value obtained in
Ref.~\cite{2020A&A...636A..17P} from dust polarization maps, but the
longitude differs by $\sim 40^\circ$. The model parameters of this fit
are close to the parameters of the \modelBase model, except for a 25\%
smaller ``striation'' factor (related to the contribution of anisotropic
random fields to \PI) and a different shape parameter of the
poloidal X-field ($p=1.0$ instead of $1.4$).

\begin{figure}[t]
  \input{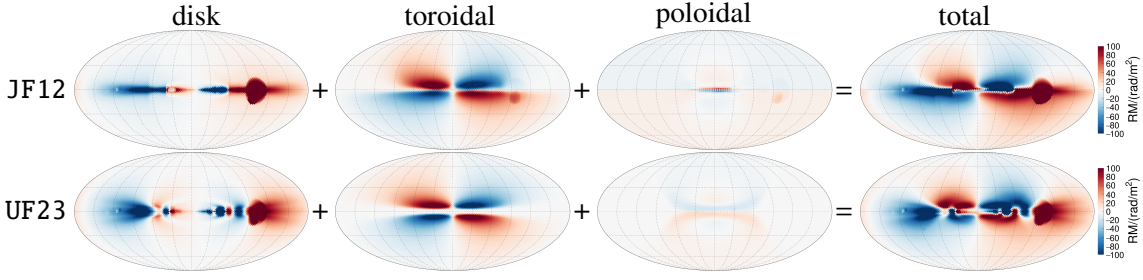}
\caption{Sky maps of predicted Rotation Measures of \JF (top row) and \UF \modelNe (bottom row). The first three columns
  show the contributions of the disk, toroidal and poloidal model components and the right-most column shows their sum.}
\label{fig:modelRMs}
\vspace{-.2in}
\end{figure}

Furthermore, we investigated in detail the differences in the toroidal
halo field of \JF and the \UF models, to identify the main reason for
the discrepancies in the magnification maps reported in
Ref.~\cite{Bister:2024ocm}. Apart from the \texttt{twistX} variant,
all the models use the same functional form for the toroidal
field. However, the \JF model, which lacked the flexible spiral arm
geometry of the UF23 models, required a significantly larger radial
extent in the Galactic South than in the North.  Both models have
azimuthal fields of opposite direction in the Northern and Southern
hemispheres, as needed to describe the observed anti-symmetric pattern
in longitude $\ell$ and latitude $b$ of rotation measures. However,
the radial extent of the toroidal field in the two hemispheres was
found to be the same in the \UF model fit ($\rT=10.1$~kpc, with a
transition width of 1.1~kpc in \modelNe).  In the case of the \JF
model, the radial extent differs in the Galactic North and South
($\rNT=9.2$~kpc $\rST=17.5$~kpc with a sharp transition width of
0.14~kpc); the different radial extents of the toroidal halos directly
leads to the ``butterfly pattern'' of the observed rotation measures,
$\text{RM}(\ell,b) \approx
\mathcircled{\!\!\frac{\scalebox{0.5}{\,\,--\,\,}\bm{+-}\scalebox{0.5}{\,\,+\,\,}}{\bm{--++}}\!\!}$,
i.e.\ to the smaller extent in longitude of the central anti-symmetric
RM features in the North than in the South. In \UF, on the other hand,
this pattern is achieved by the superposition of a halo with a
symmetric geometry and the field of the local spiral arm, similar
to~\cite{Pshirkov:2011um}. This important difference is best
appreciated by comparing how the individual model components add up to
the full \RM, as shown in Fig.~\ref{fig:modelRMs} (see e.g.\ Fig.~2(a)
in~\cite{Unger:2023lob} for the \RM pattern in data).

Since the \JF model does not describe the contemporary \RM and
synchrotron data well ($\Delta\chi^2 \approx + 1000$ with respect to
\UF), it is difficult to assess whether its lack of N-S symmetry in the radial extent of the toroidal fields is
a viable alternative to the one of the \UF models. We therefore
refitted the \UF model forcing the fit to converge to the local
minimum of an asymmetric halo geometry, made possible by fixing the toroidal width to
0.14~kpc. The resulting forced model variant has $\Delta\chi^2 =
+171$ with respect to \modelNe and $\Delta\chi^2 = +244$ with respect
to \texttt{base} when using the YWM16 thermal electron model.
While a N-S symmetric toroidal halo component is appealing, being a natural result of the radial and
vertical shear of the poloidal field as in the \texttt{twistX}
variant, the $\chi^2$ difference between the models is in the
ballpark of the differences of \UF model variants, see Table~2
in~\cite{Unger:2023lob}. Therefore, we cannot exclude the possibility
of an asymmetric halo with high confidence based on the extragalactic
\RMs alone. Note that the analysis of pulsar \RMs in Ref.~\cite{2024ApJ...966..240X} favors a North-South symmetric halo, but
relies to a large extent on the validity of pulsar distance estimates
based on their dispersion measures and assumed structure of the thermal electron distribution.

For future reference, if a model with a \JF-type asymmetric halo is
required that fits the contemporary RM and synchrotron data, we list
here the parameters of the \modelAsymT model tuned with the NE2001
thermal electron model, {\footnotesize $ (\pitch,$ $\fsz,$ $\fswz,$ $B_1,$ $B_2 ,$
$B_2,$ $\phi_1,$ $\phi_2,$ $\phi_3,$ $\BNT,$ $\BST,$ $\zT,$ $\rNT,$
$\rST,$ $\wT,$ $\BP,$ $ p,$ $ \zP,$ $\rP,$ $\wP,$ $\xi) =$ $( 12.1,$ $
0.6,$ $ 0.01,$ $1.57,$ $1.5,$ $3.5,$ $ 180,$ $ 156,$ $ 69,$
$2.28,-2.08,$ $3.6,$ $9.0,$ $ 15 (>11),$ $0.14,$ $0.94,$ $1.77,$
$3.34,$ $7.45,$ $0.17,$ $0.41)$}, with the usual units kpc, $\upmu$G and
degree, see~\cite{Unger:2023lob} for a detailed description of these
parameters.

\section{UHECR Deflections}

\begin{figure}[t]
  \input{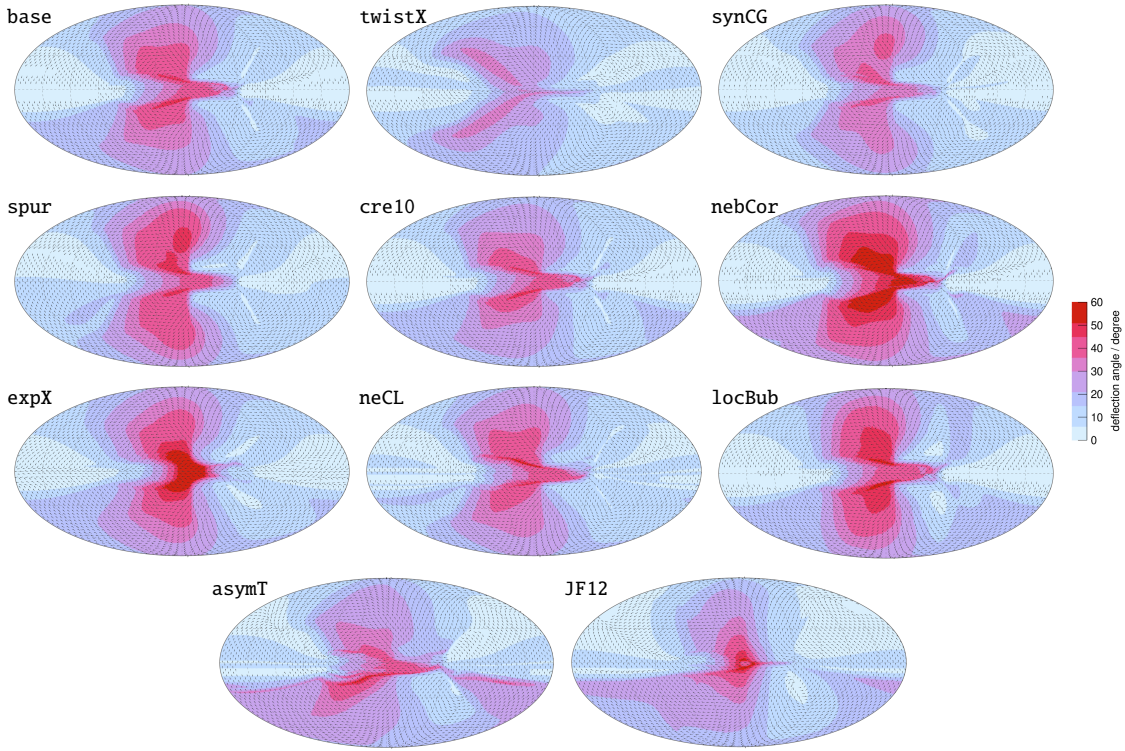}
   \caption{Angular deflections of ultrahigh-energy cosmic rays in the
     eight \UF model variations, the preliminary fit with a local bubble presented at this
     conference, \modelAsymT model and \JF. Colors and arrows denote the size and direction of
     the deflection in the GMF following the particles from Earth to
     the edge of the Galaxy. Positions on the skymap denote arrival
     directions at Earth. The rigidity is
     $2\times10^{19}$~V.} \label{fig:defl}
\end{figure}

Each of the \UF model variations introduced in the previous
section is a viable description of the GMF given the
current \RM and \PI data, and the differences between the models
gives an estimate of the lower limit on the uncertainty
of our knowledge of the magnetic field of the Galaxy.

In Fig.~\ref{fig:defl} we show sky maps of the {\itshape deflection
  angle} at a particle rigidity of 20~EV ($1~\text{EV} =
10^{18}~\text{V}$) for each of the eight \UF models, \modelAsymT, the
preliminary fit with a local bubble, and the \JF model. These were
obtained by backtracking to a galactocentric radius of
$r_\text{max}=30$~kpc. The magnitude of the deflection angle is
indicated by colors at each arrival direction at Earth on a \HEALPix
grid with $N_\text{side}=64$. The direction of the deflection is
indicated by an arrow for a subset of directions on a
$N_\text{side}=16$ grid if $\theta_\text{def} > 1^\circ$. As can be
seen, at a rigidity of $\R = 20~EV$ all models, including the
\modelAsymT and the preliminary \modelBub fit, exhibit qualitatively
similar deflection patterns but with quantitative differences for each
model.  All models exhibit a left-right asymmetry with deflections
being larger if the particle is back-tracked towards positive
longitudes and smaller for negative longitudes. This is the
consequence of the twisted nature of the halo field.

For a closer look at the differences between the models, we compare
the back-tracked directions in Fig.~\ref{fig:defl2} for a sampling of
arrival directions. The particle rigidity is again 20~EV and the lines
interpolate back-tracked directions at higher rigidities.  This figure
illustrates the similarity of the models, since in many directions all
of them roughly agree on the overall direction of the deflection, but
also shows the model uncertainties, visible as a scatter in predicted
directions for the ensemble of models.
\def\ww{0.75}
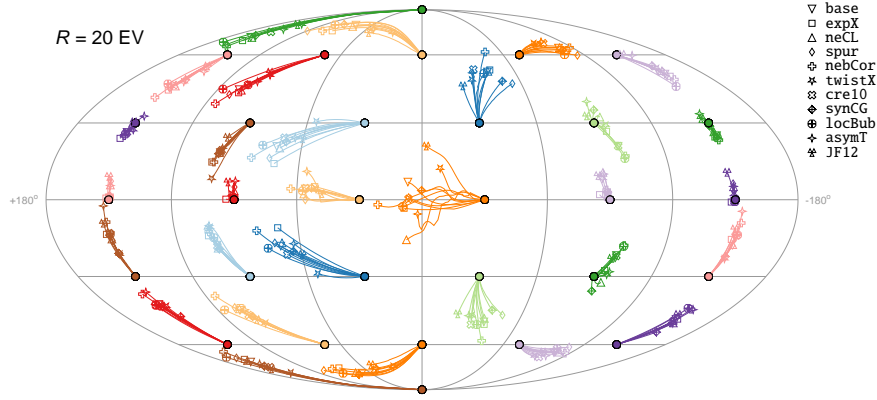
\begin{figure}[t]
  \centering
 \begin{overpic}[width=\ww\textwidth]{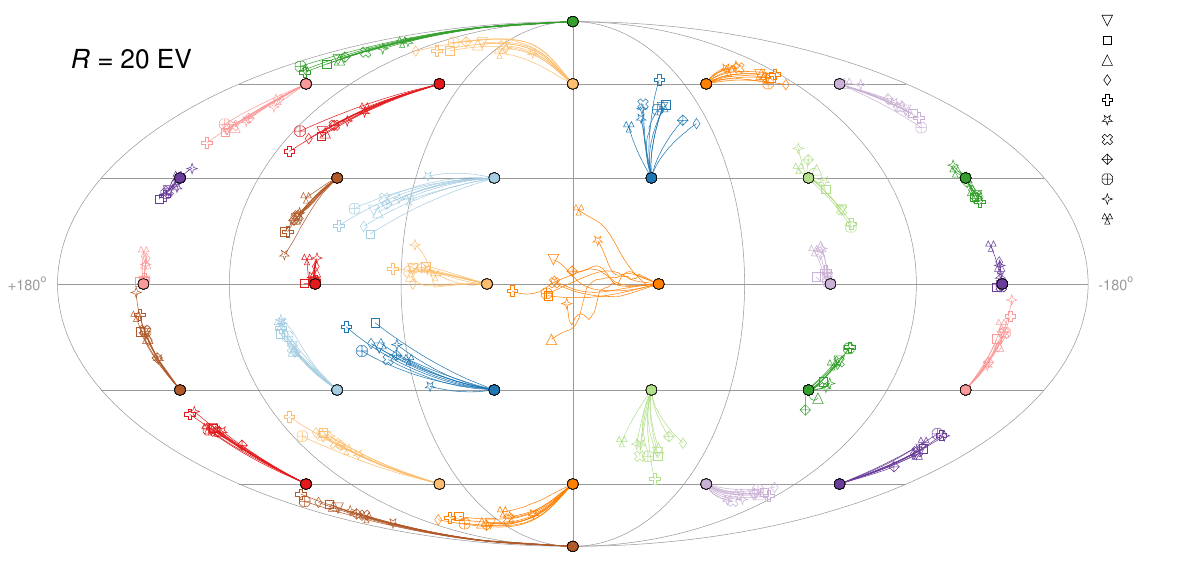}
    \def\xlab{95.3}
    \put(\xlab,45.90000){\scalebox{\ww}{\scriptsize\modelBase}}
    \put(\xlab,44.23){\scalebox{\ww}{\scriptsize\modelXr}}
    \put(\xlab,42.56){\scalebox{\ww}{\scriptsize\modelNe}}
    \put(\xlab,40.89){\scalebox{\ww}{\scriptsize\modelSpur}}
    \put(\xlab,39.22){\scalebox{\ww}{\scriptsize\modelKappa}}
    \put(\xlab,37.55){\scalebox{\ww}{\scriptsize\modelTwist}}
    \put(\xlab,35.88){\scalebox{\ww}{\scriptsize\modelCre}}
    \put(\xlab,34.21){\scalebox{\ww}{\scriptsize\modelSyn}}
    \put(\xlab,32.54){\scalebox{\ww}{\scriptsize\modelBub}}
   \put(\xlab,30.87){\scalebox{\ww}{\scriptsize\modelAsymT}}
    \put(\xlab,29.15){\scalebox{\ww}{\scriptsize\JF}}
 \end{overpic}
   \caption{Angular deflections of ultrahigh-energy cosmic rays in the
     eight \UF model variations, \JF and the preliminary fit with a
     local bubble presented at this conference. The cosmic-ray
     rigidity is $2\times10^{19}$~V. Filled circles denote
     a grid of arrival directions and the open symbols are the
     back-tracked directions at the edge of the
     Galaxy.} \label{fig:defl2}\vspace*{-0.3cm}
\end{figure}
It is worth noting that the deflections predicted by the widely-used
\JF model are generally within the range of deflections predicted for
the \UF GMF models. This is not the case for the deflections
calculated with the GMF model of \cite{Pshirkov:2011um} (not shown),
due to the absence of a poloidal component in that model. The model
components of the more recent \KST model include a toroidal field with
unequal North/South radial limits, very similar to the one of \JF and
\modelAsymT. The large magnetic field strength of the disk field of
the \KST model in the outer Galaxy is driven by the polarized
intensity from the ``Fan Region'' -- a feature in the synchrotron sky
that is usually masked in fits of the global structure of the GMF. The
magnetic field strength in this region is fully degenerate with the
assumed local cosmic-ray electron density. Furthermore, even
though the Fan Region is a large-scale feature in the outer
Galaxy~\cite{2017MNRAS.467.4631H}, it seems implausible that it is
representative for the whole Perseus spiral arm, as assumed in the
\KST model. Therefore, we do not include this model in our
evaluation of deflection uncertainties.

\section{Localization of UHECR Events}
Recently, we studied the localization of the ``Amaterasu Particle'',
the highest-energy event detected by the TA
collaboration~\citep{TAScience}, using the \UF model variations of
the coherent Galactic magnetic field to backtrack its arrival
direction at the edge of the Galaxy~\citep{Unger:2023hnu}.
For this conference we apply the same analysis to the four
highest-energy events from the Auger phase I data
set~\citep{PierreAuger:2022qcg}.
These are listed in
Table~\ref{tab:events}. The highest-energy event detected by the Pierre Auger Observatory so far has
an energy of 166~EeV.  For comparison, the nominal energy of the
Amaterasu event is $E =244\pm 29 (\text{stat.})^{+51}_{-76} \,
(\text{syst.})$~EeV using the TA energy scale. This value can be
converted to the Auger energy scale using the energy-dependent
TA-Auger relative cross-calibration factor estimated by the UHE
spectrum working group in the common declination band, $\Delta
E_\text{TA}/E_\text{TA} = -0.09 -
0.2\,(\lg(E_\text{TA}/\text{eV})-19)$~\citep{PierreAuger:2023uif}.
The correction amounts to $-36.7$\% and the energy of the Amaterasu
particle at the Auger energy scale is $154 {\pm} 18\,
(\text{stat.})$~EeV, compatible at $1.2\,\sigma_\text{syst}$ with the
nominal Amaterasu energy within the negative systematic uncertainty of
$\sigma_\text{syst} = -76$~EeV.  Note that this cross-calibration was
obtained by the assumption that the cosmic-ray energy spectra in the
Northern and Southern hemisphere are identical, which might not hold
at UHE (see Refs.~\cite{PierreAuger:2024hrj, TelescopeArray:2024tbi}
for further discussion).

\begin{wrapfigure}{r}{0.5\textwidth}
  \vspace*{-0.5cm}
  \centering
  \includegraphics[width=\linewidth]{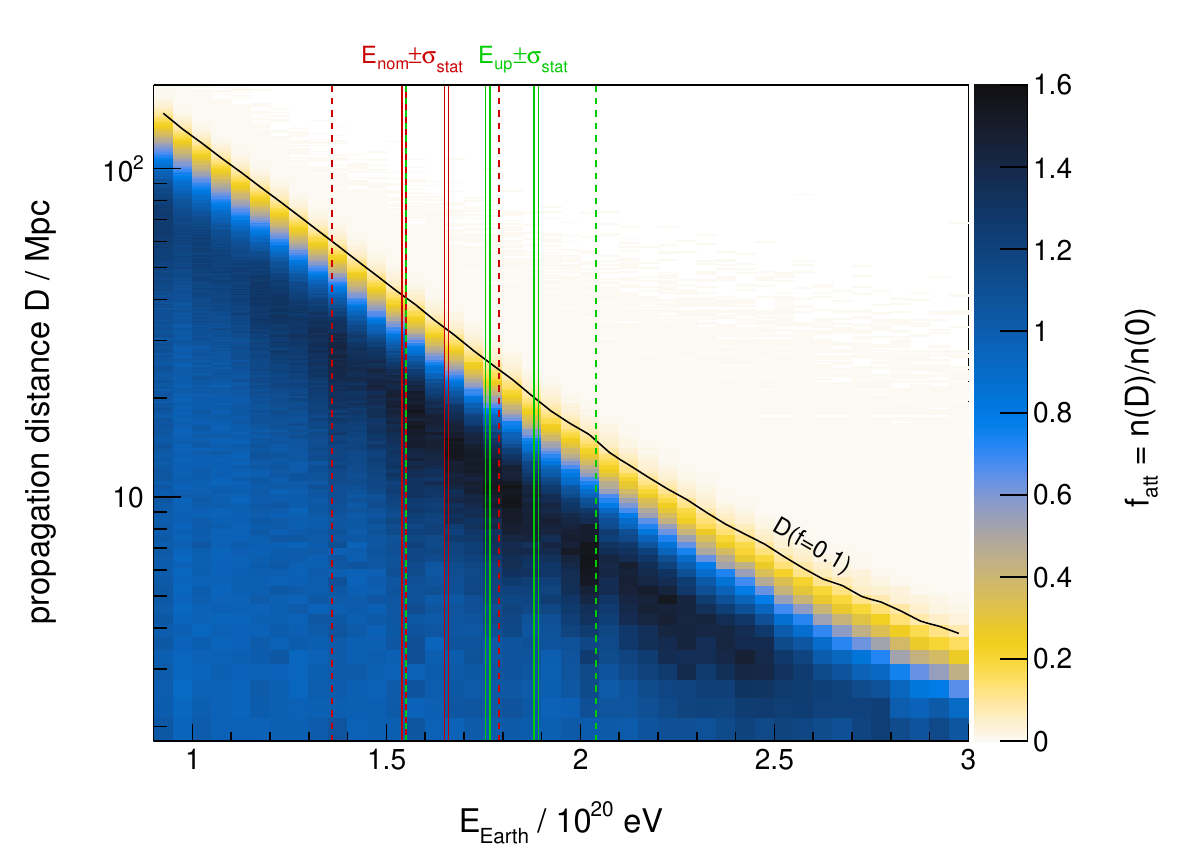}
  \caption{ Propagation distances of CRs arriving at Earth, in each
    bin of energy as given on the $x$-axis of the plot, for iron
    injected with an $\R_\text{max} = 10^{18.6}$~V energy
    spectrum (see text for details) \label{fig:distDist}}
  \vspace*{-0.3cm}
\end{wrapfigure}

\begin{table}[!t]
  \centering
  \footnotesize
   \caption{The five highest-energy cosmic-ray particles detected with
    contemporary observatories. The quoted energies are at the Auger
    energy scale and the errors are statistical. See
    text for the localization uncertainty listed in the last two columns.
    \label{tab:events}}
 \begin{tabular}{lcccccc|cc}
    id & $E$ & $\sigma_\text{stat.}$ & R.A. & \multicolumn{1}{c}{Dec.} & \multicolumn{1}{c}{$\ell$} & \multicolumn{1}{c|}{$b$} & $\Omega_\text{loc}$ / $4\uppi$ & $\theta_\text{loc}$\\
       & (EeV) & (EeV) & (degree) & \multicolumn{1}{c}{(degree)} & \multicolumn{1}{c}{(degree)} & \multicolumn{1}{c|}{(degree)} & -- &  (degree) \\ \hline
PAO191110 &166 &13  &128.9 &$-$52.0 & 269.1 & \phantom{$4$}$-6.8$ & 7.1\% & 31\\
PAO070114 &165 &13  &192.9 &$-$21.2 & 303.0 & \phantom{$-$}$41.7$ & 2.4\% & 18\\
PAO141021 &155 &12  &102.9 &$-$37.8 & 247.6 & $-16.2$ & 6.3\% & 29\\
PAO200611 &155 &12  &107.2 &$-$47.6 & 258.3 & $-16.9$& 6.6\% & 29\\
TA210527 & 154 & 18 &255.9 &\phantom{$-$}16.1  &  \phantom{$2$}36.2 & \phantom{$-$}$30.9$& 4.7\%& 25 \\
  \end{tabular}
\end{table}
Figure~\ref{fig:distDist} shows a histogram of the propagation
distance as a function of the UHECR energy at Earth, for
$\R_\text{max} = 10^{18.6}$~V. These were obtained with {\scshape
  CRPropa3}~\citep{AlvesBatista:2016vpy} assuming iron nuclei, see
\citep{Unger:2023hnu} for further details. The vertical solid lines mark the
central energy values of the four events
 for the nominal and upper ($+14\%$) Auger energy scale (note that two
 events at 155 EeV overlap, hence only four lines are visible). The
 one-standard-deviation of the statistical reconstruction uncertainty
 of the lowest- and highest-energy events is shown with dashed lines.
 We define the approximate edge of the source volume as the distance,
 $D_{0.1}$, at which the flux is attenuated by a factor of 10 relative to
 the case with no photo-interaction energy losses.  At the average
 energy of the events of 160 EeV, the maximum source distance defined
 in this way is around 35~Mpc. At the one-sigma upper Auger energy
 scale, it is 23~Mpc. An even larger range of allowed distances is
 within one standard deviation of the statistical uncertainty of the
 energy reconstructions. In the following, we will concentrate on the
 results obtained at the nominal energy scale. We checked that the
 results at a $\pm1\,\sigma_\text{syst.}$ lead to similar conclusions.

We backtracked the arrival directions of the five highest energy
cosmic-ray events through the Galactic magnetic field to identify the
domain of highest source likelihood marginalizing over statistical
rigidity uncertainty, coherent magnetic field models and different
realizations of the random magnetic field of the Galaxy. The latter are generated with a root-mean-square field
strength as given by the Planck-tune of the \JF random
field~\citep{Jansson:2012rt,Planck:2016gdp} and we adopt an outer
scale of the Kolmogorov turbulence of 100~pc, or equivalently a
coherence length of 20~pc. We assumed that all particles were iron
nuclei from the UHE end of the Peters cycle, see \citep{Unger:2023hnu}
for further discussion. In total, we backtracked 50 particles through
180 random field realizations per coherent field model, i.e.\ 9k
particles per model and thus 72k particles per event.
We then define as the {\itshape localization contour} of the particle the
backtracked directions at which the probability density drops below a
factor 0.05 of the peak value of any model (see Eq.~(2)
in~\cite{Unger:2023hnu}). This contour encloses about 95\% of all
simulated particles (which is a numerical coincidence since in general
contour levels do not correspond to confidence intervals).

The angular size of the corresponding event localization is listed in
the two last columns of Table \ref{tab:events}. Here
$\Omega_\text{loc}$ refers to the solid angle covered by the contour
and the corresponding angular localization radius, $\theta_\text{loc}
\equiv \arccos\left(1-\Omega_\text{loc}/2\uppi\right)$, contains the
same area if the localization contour were circular.
\def\figw{1}
\begin{figure}[t] \centering
  \includegraphics[clip,rviewport=-0.085 0 1.02 1,width=\figw\linewidth]{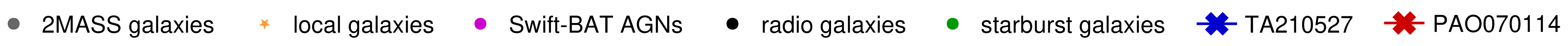}\\
  \includegraphics[width=\figw\linewidth]{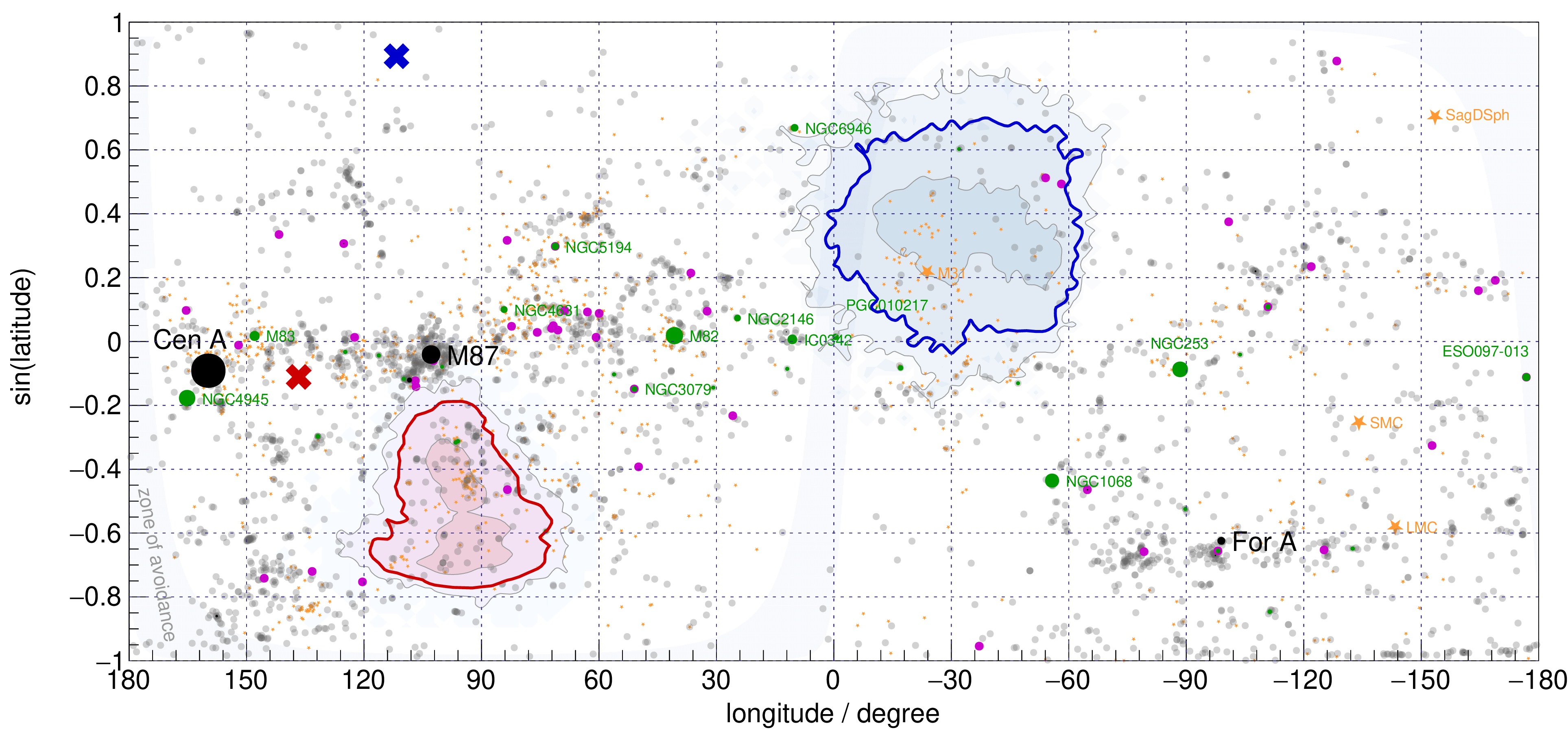}
  \caption{Contour lines of event localization of TA210527 (blue) and
    PAO070114 (red) in supergalactic coordinates.  The gray lines are
    at contour levels of 0.01 and 0.5 and in thick blue/red line at
    $0.05$. Points denote galaxies in the local volume ($D<11$ Mpc,
    orange) and galaxies within 35~Mpc from the 2MASS survey (gray),
    AGN (violet), radio galaxies (black) and starburst galaxies
    (green). The detected arrival direction of the particle is shown
    as a cross; the zone around the Galactic plane not covered in the
    2MASS galaxy survey due to dust obscuration (``zone of
    avoidance'') is indicated with a very light gray
    area.  \label{fig:localisg}}
  \vspace*{-0.2cm}
\end{figure}
Note that these
values are at the nominal Auger energy scale. The numerical value of
the localization of TA\-210527 is somewhat larger
in~\cite{Unger:2023hnu} (6.6\% instead of 4.7\%) since it includes the
uncertainty in the TA energy scale.

 The two events with the smallest localization uncertainty, TA\-210527
 and PAO\-070114, are shown in Fig.~\ref{fig:localisg} in
 supergalactic coordinates~\cite{2000MNRAS.312..166L}.  Due to their
 arrival directions far from the Galactic plane, their paths avoid
 regions of high magnetic field strength. However, the trajectories of
 PAO\-191110, PAO\-141021 and PAO\-200611 pass near the Galactic plane
 where the realization-to-realization variance of the random GMF is
 large. We stress that the deflections depend significantly on the random field, therefore the angular size of the localization regions can be expected to change to some degree
when a future improved version of the coherent plus random GMF is
 developed, in which the random field for each of the suite of eight
 coherent GMF models is self-consistently fit to the synchrotron total
 intensity. The localization regions shown in   Fig. \ref{fig:localisg} include variations across the eight coherent fields of \UF but have not been extended to include the two new models \modelAsymT and
 \modelBub, introduced in these proceedings.  However, the central deflection for these models falls well within the contours shown, except that the preliminary \modelBub
 model predicts a central direction for PAO\-070114 that is slightly outside the
 contour at its lower left edge. However, this deviation does not qualitatively affect the subsequent discussion.

 Fig. \ref{fig:localisg} also shows the position of
galaxies within 35 Mpc, from the catalogs
a) the flux-limited Two Micron All-Sky galaxy
survey~\citep{Huchra:2011ii} cross-matched with the HyperLEDA distance
database~\citep{Makarov:2014txa} (``2MASS" herein), b) the flux-limited
\textit{Swift}-BAT 105-month catalog of active galactic nuclei (AGN)
observed in hard X-rays \citep{2018ApJS..235....4O} and c) a sample of
nearby starburst galaxies \cite{2019JCAP...10..073L}. We also investigate d) radio galaxies from the volume-limited
catalog~\cite{vanVelzen:2012fn}. Not all radio galaxies pass the
UHE luminosity criterion~\citep{Waxman:1995vg,
  Blandford:1999hi,Farrar:2008ex} to be viable candidates for
accelerators up to the rigidity of the Amaterasu particle and we
therefore use only the subset identified as satisfying that
requirement by \cite{Matthews:2018laz}.  For completeness, we also
show galaxies from the "local volume catalog"
\cite{2018MNRAS.479.4136K}, a volume-limited sample of galaxies within 11~Mpc.

PAO\-070114 is backtracked close to M87 but an origin in M87 has a low
probability given that its location is outside of the tail of the
contour, i.e.\ none of the 9000 backtracked particles per GMF model
arrived at its position. Within the assigned uncertainty, none of the
usual suspects of astrophysical candidate accelerators therefore can
be associated with this event. This conclusion is similar to the one
drawn in our previous analysis for TA's Amaterasu particle, suggesting
that both these UHECRs may originate in transient events in otherwise
undistinguished galaxies.  In addition, this analysis focuses
attention on the possibility that Swift-BAT AGNs, perhaps via
transients they host~\cite{Farrar:2008ex}, may be the principle
sources of UHECRs. All four of the highest energy Auger events contain
at least one Swift-BAT AGN in their localization volume -- albeit only
on the edge of the localization region in the cases of PAO191110 and
PAO070114.  Moreover, using the central Auger energy assignment for
Amaterasu brings a similarly peripheral AGN into its localization
region.  This does not mean that there is a Swift-BAT AGN source
candidate for every one of the highest energy events -- for that to be
established, a single GMF and realization of the random field would
have to be found with this feature for all, through a
realization-by-realization query and a more detailed analysis of the
energy-assignment-sensitivity of these conclusions.

\footnotesize
\setlength{\bibsep}{0pt plus 0.3ex}
\begin{multicols}{2}
  \setlength\columnsep{3pt}
 \setstretch{0.9}
  \bibliographystyle{uhecr}
\bibliography{main}
\end{multicols}

\end{document}